\documentclass[preprint,showpacs,pra]{revtex4}
\usepackage{amssymb}
\usepackage{amsmath}
\usepackage{graphics}
\usepackage{color}

\definecolor{red}{rgb}{1,0,0}
\input{tcilatex}
\begin{document}

\title{Shell-to-shell ionization cross sections of antiprotons, H$^{+}$,\\
He$^{2+}$, Be$^{4+}$, C$^{6+}$ and O$^{8+}$ on H, \ He, Li,\ Be, B, C, N, O,
F, Ne, P, S and Ar neutral atoms}
\author{J. E. Miraglia}
\date{\today }
\affiliation{Instituto de Astronom\'{\i}a y F\'{\i}sica del Espacio. Consejo Nacional de
Investigaciones Cient\'{\i}ficas y T\'{e}cnicas}
\affiliation{Departamento de F\'{\i}sica. Facultad de Ciencias Exactas y Naturales.
Universidad de Buenos Aires. \\
Casilla de Correo 67, Sucursal 28, {C1428EGA} Buenos Aires, Argentina.}

\begin{abstract}
Total ionization cross sections of H, \ He, Li,\ Be, B, C, N, O, F, Ne, P S
and Ar neutral atoms by impact of antiprotons, H$^{+}$, He$^{2+}$, Be$^{4+}$%
, C$^{6+}$ and O$^{8+}$. were calculated using the CDWEIS (continuum
distorted wave -Eikonal Initial state) theoretical method. Cross section
depending on of initial the quantum numbers $n$ and $l$ are reported in
Tables for a range of impact energies covering from 100 keV/amu to 10 MeV/amu
\end{abstract}

\pacs{ }
\maketitle

Total ionization cross sections of\ H, \ He, Li,\ Be, B, C, N, O, F, Ne, P,
S and Ar neutral atoms by impact of $\overline{p}$ (antiprotons), H$^{+}$, He%
$^{2+}$, Be$^{4+}$, C$^{6+}$ and O$^{8+}$. were calculated using the CDW-EIS
(continuum distorted wave -Eikonal Initial state) theoretical method. The
total ionization cross section of an electron in the $nlm$ initial state,
due to the interaction with a projectile of charge $Z_{P}$ and impact
velocity $v$, is given by the four-dimensional integral

\begin{equation}
\sigma _{nl}=\sum_{m=.l}^{l}\frac{(2\pi )^{2}}{v^{2}}\int d\overrightarrow{E}%
\int d\overrightarrow{\eta }\ \left\vert T_{\overrightarrow{k},nlm}(%
\overrightarrow{E},\overrightarrow{\eta })\right\vert ^{2},  \label{1}
\end{equation}%
where $T_{\overrightarrow{k},nlm}(\overrightarrow{E},\overrightarrow{\eta })$
is the transition matrix as a function of is the component of the momentum
transfer $\overrightarrow{\eta }$ perpendicular to the incident velocity $%
\vec{v}.\ $In our theoretical treatment we expand our final continuum wave
function on the target atom in the usual form

\begin{equation}
\psi _{\overrightarrow{k}}^{-}(\overrightarrow{r})=\sum_{l=0}^{l_{\max
}}\sum_{m=-l}^{l}R_{kl}^{-}(r)Y_{l}^{m}(\widehat{r})Y_{l}^{m^{\ast }}(%
\widehat{k}),  \label{2}
\end{equation}%
We are confident with our calculations up to $l_{\max }\sim 30.$ As the
impact velocity $v$\ increases we would require of larger $l_{\max }$ in (1)$%
.$ At the highest impact energies here reported we estimate a deficiency of
our results of about 2-3\%.\ The wave functions of initial bound state
characterized by the quantum numbers $nlm\ $and final continuum state $\psi
_{\overrightarrow{k}}^{-}(\overrightarrow{r})$were obtained by using the
RADIALF code developed by Salvat and co-workers using a Hartree-Fock
potential obtained the Depurated Inversion Model \cite{mendez2016,mendez2018}
. Details of the calculation can be seen in \cite{montanari2017}, and for
proton in Refs.\cite{miraglia2008,miraglia2009} \ In parallel to this
article, we present ionization cross sections of biomolecules of interest by
using these results along with the stoichiometrical approximation \cite%
{miraglia2019}. Note we complete the first and second rows of the periodic
table, and with P and S and Ar we cover three\ elements of the third row.

The Tables that follows reports the calculated cross section $\sigma _{nl}\ $%
in atomic units

\newpage

\begin{table}[tbph]
\caption{Total ionization cross sections for multicharge bare ions on
Hydrogen at different impact energies. Cross sections are in atomic units
divided $Z_{P}^{2}$, and projectile energies are in MeV/amu. To save space,
throughout these tables the subindex $n$ replaces the $10^{n}$ factor.}
\label{table1}%
\begin{tabular}{|r|c|llllllllllll|}
\hline\hline
$Z_P$\  & $\ nl\ \ $ & \ 0.1\  & 0.2 & 0.3 & 0.4 & 0.5 & 0.7 & 1 & 2 & 3 & 5
& 7 & 10 \\ \hline
-1 & 1s & 3.30$_{ }$ & 2.33$_{ }$ & 1.77$_{ }$ & 1.43$_{ }$ & 1.20$_{ }$ & 
9.11$_{-1}$ & 6.75$_{-1}$ & 3.71$_{-1}$ & 2.59$_{-1}$ & 1.64$_{-1}$ & 1.22$%
_{-1}$ & 8.82$_{-2}$ \\ 
1 & 1s & 4.24$_{ }$ & 2.64$_{ }$ & 1.91$_{ }$ & 1.50$_{ }$ & 1.25$_{ }$ & 
9.34$_{-1}$ & 6.86$_{-1}$ & 3.73$_{-1}$ & 2.69$_{-1}$ & 1.70$_{-1}$ & 1.26$%
_{-1}$ & 8.82$_{-2}$ \\ 
2 & 1s & 3.13$_{ }$ & 2.33$_{ }$ & 1.77$_{ }$ & 1.43$_{ }$ & 1.20$_{ }$ & 
9.09$_{-1}$ & 6.74$_{-1}$ & 3.70$_{-1}$ & 2.67$_{-1}$ & 1.69$_{-1}$ & 1.25$%
_{-1}$ & 8.81$_{-2}$ \\ 
4 & 1s & 1.66$_{ }$ & 1.67$_{ }$ & 1.41$_{ }$ & 1.20$_{ }$ & 1.04$_{ }$ & 
8.24$_{-1}$ & 6.29$_{-1}$ & 3.58$_{-1}$ & 2.53$_{-1}$ & 1.67$_{-1}$ & 1.24$%
_{-1}$ & 8.76$_{-2}$ \\ 
6 & 1s & 9.39$_{-1}$ & 1.22$_{ }$ & 1.12$_{ }$ & 1.00$_{ }$ & 8.94$_{-1}$ & 
7.33$_{-1}$ & 5.76$_{-1}$ & 3.41$_{-1}$ & 2.53$_{-1}$ & 1.64$_{-1}$ & 1.23$%
_{-1}$ & 8.67$_{-2}$ \\ 
8 & 1s & 5.60$_{-1}$ & 9.08$_{-1}$ & 9.05$_{-1}$ & 8.42$_{-1}$ & 7.72$_{-1}$
& 6.54$_{-1}$ & 5.28$_{-1}$ & 3.23$_{-1}$ & 2.36$_{-1}$ & 1.60$_{-1}$ & 1.20$%
_{-1}$ & 8.55$_{-2}$ \\ \hline\hline
\end{tabular}%
\end{table}

\begin{table}[htbp]
\caption{Total ionization cross sections for multicharge bare ions on Helium
at different impact energies. Cross sections are in atomic units divided $%
Z_P^2$, and projectile energies are in MeV/amu. To save space, throughout
these tables the subindex $n$ replaces the $10^n$ factor.}%
\begin{tabular}{|r|c|llllllllllll|}
\hline\hline
$Z_P$\  & $\ nl\ \ $ & \ 0.1\  & 0.2 & 0.3 & 0.4 & 0.5 & 0.7 & 1 & 2 & 3 & 5
& 7 & 10 \\ \hline
-1 & 1s & 2.25$_{ }$ & 1.96$_{ }$ & 1.63$_{ }$ & 1.39$_{ }$ & 1.21$_{ }$ & 
9.60$_{-1}$ & 7.38$_{-1}$ & 4.27$_{-1}$ & 3.05$_{-1}$ & 1.98$_{-1}$ & 1.49$%
_{-1}$ & 1.09$_{-1}$ \\ 
1 & 1s & 3.17$_{ }$ & 2.45$_{ }$ & 1.89$_{ }$ & 1.54$_{ }$ & 1.31$_{ }$ & 
1.01$_{ }$ & 7.60$_{-1}$ & 4.31$_{-1}$ & 3.07$_{-1}$ & 1.99$_{-1}$ & 1.49$%
_{-1}$ & 1.09$_{-1}$ \\ 
2 & 1s & 2.00$_{ }$ & 2.03$_{ }$ & 1.70$_{ }$ & 1.43$_{ }$ & 1.24$_{ }$ & 
9.73$_{-1}$ & 7.42$_{-1}$ & 4.27$_{-1}$ & 3.05$_{-1}$ & 1.98$_{-1}$ & 1.48$%
_{-1}$ & 1.09$_{-1}$ \\ 
4 & 1s & 7.77$_{-1}$ & 1.25$_{ }$ & 1.22$_{ }$ & 1.11$_{ }$ & 1.01$_{ }$ & 
8.40$_{-1}$ & 6.71$_{-1}$ & 4.07$_{-1}$ & 2.96$_{-1}$ & 1.95$_{-1}$ & 1.47$%
_{-1}$ & 1.08$_{-1}$ \\ 
6 & 1s & 3.34$_{-1}$ & 8.02$_{-1}$ & 8.82$_{-1}$ & 8.57$_{-1}$ & 8.08$_{-1}$
& 7.07$_{-1}$ & 5.89$_{-1}$ & 3.79$_{-1}$ & 2.82$_{-1}$ & 1.89$_{-1}$ & 1.44$%
_{-1}$ & 1.07$_{-1}$ \\ 
8 & 1s & 1.53$_{-1}$ & 5.40$_{-1}$ & 6.60$_{-1}$ & 6.75$_{-1}$ & 6.57$_{-1}$
& 6.00$_{-1}$ & 5.17$_{-1}$ & 3.50$_{-1}$ & 2.67$_{-1}$ & 1.82$_{-1}$ & 1.40$%
_{-1}$ & 1.05$_{-1}$ \\ \hline\hline
\end{tabular}%
\end{table}

\newpage

\begin{table}[htbp]
\caption{Total ionization cross sections for multicharge bare ions on
Lithium at different impact energies. Cross sections are in atomic units
divided $Z_P^2$, and projectile energies are in MeV/amu}%
\begin{tabular}{|r|r|llllllllllll|}
\hline\hline
$Z_P$ & $nl$ & \ 0.1 & 0.2 & 0.3 & 0.4 & 0.5 & 0.7 & 1 & 2 & 3 & 5 & 7 & 10
\\ \hline
-1 & 2s & 8.19$_{ }$ & 4.56$_{ }$ & 3.15$_{ }$ & 2.46$_{ }$ & 1.96$_{ }$ & 
1.42$_{ }$ & 1.01$_{ }$ & 5.29$_{-1}$ & 3.58$_{-1}$ & 2.15$_{-1}$ & 1.55$%
_{-1}$ & 1.12$_{-1}$ \\ 
-1 & 1s & 1.52$_{-1}$ & 1.84$_{-1}$ & 1.81$_{-1}$ & 1.69$_{-1}$ & 1.57$_{-1}$
& 1.34$_{-1}$ & 1.09$_{-1}$ & 6.73$_{-2}$ & 4.93$_{-2}$ & 3.27$_{-2}$ & 2.48$%
_{-2}$ & 1.83$_{-2}$ \\ 
-1 & Total & 8.50$_{ }$ & 4.92$_{ }$ & 3.51$_{ }$ & 2.79$_{ }$ & 2.27$_{ }$
& 1.69$_{ }$ & 1.23$_{ }$ & 6.64$_{-1}$ & 4.56$_{-1}$ & 2.81$_{-1}$ & 2.05$%
_{-1}$ & 1.49$_{-1}$ \\ \hline
1 & 2s & 8.79$_{ }$ & 4.72$_{ }$ & 3.22$_{ }$ & 2.49$_{ }$ & 1.98$_{ }$ & 
1.43$_{ }$ & 1.02$_{ }$ & 5.29$_{-1}$ & 3.58$_{-1}$ & 2.19$_{-1}$ & 1.55$%
_{-1}$ & 1.12$_{-1}$ \\ 
1 & 1s & 2.21$_{-1}$ & 2.73$_{-1}$ & 2.46$_{-1}$ & 2.14$_{-1}$ & 1.89$_{-1}$
& 1.52$_{-1}$ & 1.18$_{-1}$ & 6.93$_{-2}$ & 5.01$_{-2}$ & 3.29$_{-2}$ & 2.49$%
_{-2}$ & 1.84$_{-2}$ \\ 
1 & Total & 9.23$_{ }$ & 5.26$_{ }$ & 3.71$_{ }$ & 2.92$_{ }$ & 2.36$_{ }$ & 
1.74$_{ }$ & 1.25$_{ }$ & 6.68$_{-1}$ & 4.58$_{-1}$ & 2.84$_{-1}$ & 2.05$%
_{-1}$ & 1.49$_{-1}$ \\ \hline
2 & 2s & 7.54$_{ }$ & 4.45$_{ }$ & 3.12$_{ }$ & 2.44$_{ }$ & 1.95$_{ }$ & 
1.42$_{ }$ & 1.01$_{ }$ & 5.28$_{-1}$ & 3.57$_{-1}$ & 2.18$_{-1}$ & 1.55$%
_{-1}$ & 1.12$_{-1}$ \\
\hline\hline
\end{tabular}%
\end{table}

\newpage

\begin{table}[htbp]
\caption{Total ionization cross sections for multicharge bare ions on
Beryllium at different impact energies. Cross sections are in atomic units
divided $Z_P^2$, and projectile energies are in MeV/amu}%
\begin{tabular}{|r|r|llllllllllll|}
\hline\hline
$Z_P$ & $nl$ & \ 0.1 & 0.2 & 0.3 & 0.4 & 0.5 & 0.7 & 1 & 2 & 3 & 5 & 7 & 10
\\ \hline
-1 & 2s & 4.73$_{ }$ & 2.85$_{ }$ & 2.04$_{ }$ & 1.59$_{ }$ & 1.31$_{ }$ & 
9.67$_{-1}$ & 7.00$_{-1}$ & 3.74$_{-1}$ & 2.54$_{-1}$ & 1.57$_{-1}$ & 1.15$%
_{-1}$ & 8.23$_{-2}$ \\ 
-1 & 1s & 3.02$_{-2}$ & 4.77$_{-2}$ & 5.40$_{-2}$ & 5.53$_{-2}$ & 5.45$_{-2}$
& 5.07$_{-2}$ & 4.44$_{-2}$ & 2.99$_{-2}$ & 2.26$_{-2}$ & 1.53$_{-2}$ & 1.18$%
_{-2}$ & 8.80$_{-3}$ \\ 
-1 & Total & 9.52$_{ }$ & 5.80$_{ }$ & 4.18$_{ }$ & 3.29$_{ }$ & 2.72$_{ }$
& 2.04$_{ }$ & 1.49$_{ }$ & 8.09$_{-1}$ & 5.52$_{-1}$ & 3.46$_{-1}$ & 2.53$%
_{-1}$ & 1.82$_{-1}$ \\ \hline
1 & 2s & 5.63$_{ }$ & 3.07$_{ }$ & 2.13$_{ }$ & 1.64$_{ }$ & 1.34$_{ }$ & 
9.82$_{-1}$ & 7.06$_{-1}$ & 3.71$_{-1}$ & 2.54$_{-1}$ & 1.58$_{-1}$ & 1.15$%
_{-1}$ & 8.23$_{-2}$ \\ 
1 & 1s & 4.90$_{-2}$ & 7.02$_{-2}$ & 7.33$_{-2}$ & 7.06$_{-2}$ & 6.66$_{-2}$
& 5.83$_{-2}$ & 4.87$_{-2}$ & 3.11$_{-2}$ & 2.31$_{-2}$ & 1.55$_{-2}$ & 1.18$%
_{-2}$ & 8.83$_{-3}$ \\ 
1 & Total & 1.14$_{ 1}$ & 6.28$_{ }$ & 4.41$_{ }$ & 3.43$_{ }$ & 2.81$_{ }$
& 2.08$_{ }$ & 1.51$_{ }$ & 8.04$_{-1}$ & 5.55$_{-1}$ & 3.46$_{-1}$ & 2.54$%
_{-1}$ & 1.82$_{-1}$ \\ \hline
2 & 2s & 3.99$_{ }$ & 2.67$_{ }$ & 1.97$_{ }$ & 1.55$_{ }$ & 1.28$_{ }$ & 
9.55$_{-1}$ & 6.94$_{-1}$ & 3.73$_{-1}$ & 2.53$_{-1}$ & 1.57$_{-1}$ & 1.15$%
_{-1}$ & 8.22$_{-2}$ \\
\hline\hline
\end{tabular}%
\end{table}

\newpage

\begin{table}[htbp]
\caption{Total ionization cross sections for multicharge bare ions on Boron
at different impact energies. Cross sections are in atomic units divided $%
Z_P^2$, and projectile energies are in MeV/amu}%
\begin{tabular}{|r|r|llllllllllll|}
\hline\hline
-1 & 2p & 6.63$_{ }$ & 4.80$_{ }$ & 3.72$_{ }$ & 3.04$_{ }$ & 2.58$_{ }$ & 
1.99$_{ }$ & 1.50$_{ }$ & 8.43$_{-1}$ & 5.98$_{-1}$ & 3.85$_{-1}$ & 2.87$%
_{-1}$ & 2.10$_{-1}$ \\ 
-1 & 2s & 2.54$_{ }$ & 1.66$_{ }$ & 1.22$_{ }$ & 9.62$_{-1}$ & 7.98$_{-1}$ & 
5.97$_{-1}$ & 4.36$_{-1}$ & 2.33$_{-1}$ & 1.61$_{-1}$ & 1.00$_{-1}$ & 7.36$%
_{-2}$ & 5.29$_{-2}$ \\ 
-1 & 1s & 7.59$_{-3}$ & 1.47$_{-2}$ & 1.89$_{-2}$ & 2.10$_{-2}$ & 2.19$_{-2}$
& 2.21$_{-2}$ & 2.07$_{-2}$ & 1.55$_{-2}$ & 1.22$_{-2}$ & 8.51$_{-3}$ & 6.60$%
_{-3}$ & 4.99$_{-3}$ \\ 
-1 & Total & 1.17$_{ 1}$ & 8.14$_{ }$ & 6.19$_{ }$ & 5.01$_{ }$ & 4.22$_{ }$
& 3.23$_{ }$ & 2.41$_{ }$ & 1.34$_{ }$ & 9.44$_{-1}$ & 6.03$_{-1}$ & 4.48$%
_{-1}$ & 3.26$_{-1}$ \\ \hline
1 & 2p & 8.13$_{ }$ & 5.29$_{ }$ & 3.94$_{ }$ & 3.16$_{ }$ & 2.65$_{ }$ & 
2.03$_{ }$ & 1.51$_{ }$ & 8.46$_{-1}$ & 5.99$_{-1}$ & 3.85$_{-1}$ & 2.88$%
_{-1}$ & 2.10$_{-1}$ \\ 
1 & 2s & 2.85$_{ }$ & 1.77$_{ }$ & 1.27$_{ }$ & 9.95$_{-1}$ & 8.19$_{-1}$ & 
6.07$_{-1}$ & 4.40$_{-1}$ & 2.34$_{-1}$ & 1.61$_{-1}$ & 1.01$_{-1}$ & 7.37$%
_{-2}$ & 5.29$_{-2}$ \\ 
1 & 1s & 8.13$_{-3}$ & 2.19$_{-2}$ & 2.88$_{-2}$ & 3.11$_{-2}$ & 3.13$_{-2}$
& 2.93$_{-2}$ & 2.53$_{-2}$ & 1.70$_{-2}$ & 1.28$_{-2}$ & 8.72$_{-3}$ & 6.70$%
_{-3}$ & 5.03$_{-3}$ \\ 
1 & Total & 1.39$_{ 1}$ & 8.88$_{ }$ & 6.54$_{ }$ & 5.21$_{ }$ & 4.35$_{ }$
& 3.30$_{ }$ & 2.44$_{ }$ & 1.35$_{ }$ & 9.47$_{-1}$ & 6.04$_{-1}$ & 4.48$%
_{-1}$ & 3.26$_{-1}$ \\
\hline\hline
\end{tabular}%
\end{table}

\newpage

\begin{table}[tbph]
\caption{Total ionization cross sections for multicharge bare ions on Carbon
at different impact energies. Cross sections are in atomic units divided $%
Z_{P}^{2}$, and projectile energies are in MeV/amu}%
\begin{tabular}{|r|r|llllllllllll|}
\hline\hline
$Z_P$ & $nl$ & \ 0.1 & 0.2 & 0.3 & 0.4 & 0.5 & 0.7 & 1 & 2 & 3 & 5 & 7 & 10
\\ \hline
-1 & 2p & 3.74$_{ }$ & 2.95$_{ }$ & 2.38$_{ }$ & 1.99$_{ }$ & 1.71$_{ }$ & 
1.34$_{ }$ & 1.02$_{ }$ & 5.89$_{-1}$ & 4.21$_{-1}$ & 2.74$_{-1}$ & 2.06$%
_{-1}$ & 1.51$_{-1}$ \\ 
-1 & 2s & 1.49$_{ }$ & 1.05$_{ }$ & 7.90$_{-1}$ & 6.34$_{-1}$ & 5.30$_{-1}$
& 4.01$_{-1}$ & 2.95$_{-1}$ & 1.60$_{-1}$ & 1.11$_{-1}$ & 6.95$_{-2}$ & 5.11$%
_{-2}$ & 3.68$_{-2}$ \\ 
-1 & 1s & 2.26$_{-3}$ & 5.19$_{-3}$ & 7.33$_{-3}$ & 8.77$_{-3}$ & 9.67$_{-3}$
& 1.05$_{-2}$ & 1.06$_{-2}$ & 8.83$_{-3}$ & 7.22$_{-3}$ & 5.23$_{-3}$ & 4.11$%
_{-3}$ & 3.14$_{-3}$ \\ 
-1 & Total & 1.05$_{ 1}$ & 8.01$_{ }$ & 6.35$_{ }$ & 5.26$_{ }$ & 4.50$_{ }$
& 3.51$_{ }$ & 2.66$_{ }$ & 1.51$_{ }$ & 1.08$_{ }$ & 6.97$_{-1}$ & 5.22$%
_{-1}$ & 3.82$_{-1}$ \\ \hline
1 & 2p & 4.82$_{ }$ & 3.39$_{ }$ & 2.59$_{ }$ & 2.10$_{ }$ & 1.78$_{ }$ & 
1.38$_{ }$ & 1.04$_{ }$ & 5.92$_{-1}$ & 4.23$_{-1}$ & 2.74$_{-1}$ & 2.06$%
_{-1}$ & 1.51$_{-1}$ \\ 
1 & 2s & 1.69$_{ }$ & 1.14$_{ }$ & 8.38$_{-1}$ & 6.63$_{-1}$ & 5.49$_{-1}$ & 
4.10$_{-1}$ & 2.99$_{-1}$ & 1.61$_{-1}$ & 1.11$_{-1}$ & 6.96$_{-2}$ & 5.11$%
_{-2}$ & 3.68$_{-2}$ \\ 
1 & 1s & 2.02$_{-3}$ & 7.13$_{-3}$ & 1.09$_{-2}$ & 1.31$_{-2}$ & 1.42$_{-2}$
& 1.46$_{-2}$ & 1.36$_{-2}$ & 9.95$_{-3}$ & 7.75$_{-3}$ & 5.41$_{-3}$ & 4.20$%
_{-3}$ & 3.18$_{-3}$ \\
\hline\hline
\end{tabular}%
\end{table}

\newpage 
\begin{table}[htbp]
\caption{Total ionization cross sections for multicharge bare ions on
Nitrogen at different impact energies. Cross sections are in atomic units
divided $Z_P^2$, and projectile energies are in MeV/amu}%
\begin{tabular}{|r|r|llllllllllll|}
\hline\hline
$Z_P$ & $nl$ & \ 0.1 & 0.2 & 0.3 & 0.4 & 0.5 & 0.7 & 1 & 2 & 3 & 5 & 7 & 10
\\ \hline
-1 & 2p & 2.21$_{ }$ & 1.87$_{ }$ & 1.56$_{ }$ & 1.33$_{ }$ & 1.16$_{ }$ & 
9.23$_{-1}$ & 7.13$_{-1}$ & 4.17$_{-1}$ & 3.01$_{-1}$ & 1.97$_{-1}$ & 1.49$%
_{-1}$ & 1.10$_{-1}$ \\ 
-1 & 2s & 9.16$_{-1}$ & 6.97$_{-1}$ & 5.42$_{-1}$ & 4.42$_{-1}$ & 3.73$_{-1}$
& 2.85$_{-1}$ & 2.12$_{-1}$ & 1.16$_{-1}$ & 8.13$_{-2}$ & 5.13$_{-2}$ & 3.78$%
_{-2}$ & 2.73$_{-2}$ \\ 
-1 & 1s & 7.54$_{-4}$ & 2.00$_{-3}$ & 3.08$_{-3}$ & 3.92$_{-3}$ & 4.54$_{-3}$
& 5.29$_{-3}$ & 5.71$_{-3}$ & 5.28$_{-3}$ & 4.53$_{-3}$ & 3.41$_{-3}$ & 2.73$%
_{-3}$ & 2.11$_{-3}$ \\ 
-1 & Total & 8.46$_{ }$ & 7.02$_{ }$ & 5.77$_{ }$ & 4.88$_{ }$ & 4.23$_{ }$
& 3.35$_{ }$ & 2.58$_{ }$ & 1.50$_{ }$ & 1.07$_{ }$ & 7.01$_{-1}$ & 5.27$%
_{-1}$ & 3.88$_{-1}$ \\ \hline
1 & 2p & 2.95$_{ }$ & 2.24$_{ }$ & 1.75$_{ }$ & 1.44$_{ }$ & 1.23$_{ }$ & 
9.58$_{-1}$ & 7.29$_{-1}$ & 4.21$_{-1}$ & 3.02$_{-1}$ & 1.97$_{-1}$ & 1.49$%
_{-1}$ & 1.10$_{-1}$ \\ 
1 & 2s & 1.04$_{ }$ & 7.69$_{-1}$ & 5.82$_{-1}$ & 4.67$_{-1}$ & 3.90$_{-1}$
& 2.94$_{-1}$ & 2.16$_{-1}$ & 1.17$_{-1}$ & 8.16$_{-2}$ & 5.14$_{-2}$ & 3.79$%
_{-2}$ & 2.73$_{-2}$ \\ 
1 & 1s & 5.69$_{-4}$ & 2.52$_{-3}$ & 4.40$_{-3}$ & 5.76$_{-3}$ & 6.65$_{-3}$
& 7.48$_{-3}$ & 7.60$_{-3}$ & 6.14$_{-3}$ & 4.97$_{-3}$ & 3.57$_{-3}$ & 2.81$%
_{-3}$ & 2.15$_{-3}$ \\
\hline\hline
\end{tabular}%
\end{table}

\newpage 
\begin{table}[htbp]
\caption{Total ionization cross sections for multicharge bare ions on Oxigen
at different impact energies. Cross sections are in atomic units divided $%
Z_P^2$, and projectile energies are in MeV/amu}%
\begin{tabular}{|r|r|llllllllllll|}
\hline\hline
$Z_P$ & $nl$ & \ 0.1 & 0.2 & 0.3 & 0.4 & 0.5 & 0.7 & 1 & 2 & 3 & 5 & 7 & 10
\\ \hline
-1 & 2p & 1.61$_{ }$ & 1.41$_{ }$ & 1.19$_{ }$ & 1.02$_{ }$ & 8.99$_{-1}$ & 
7.23$_{-1}$ & 5.63$_{-1}$ & 3.33$_{-1}$ & 2.41$_{-1}$ & 1.59$_{-1}$ & 1.20$%
_{-1}$ & 8.87$_{-2}$ \\ 
-1 & 2s & 5.68$_{-1}$ & 4.68$_{-1}$ & 3.77$_{-1}$ & 3.13$_{-1}$ & 2.67$_{-1}$
& 2.06$_{-1}$ & 1.55$_{-1}$ & 8.62$_{-2}$ & 6.06$_{-2}$ & 3.84$_{-2}$ & 2.84$%
_{-2}$ & 2.06$_{-2}$ \\ 
-1 & 1s & 2.74$_{-4}$ & 8.28$_{-4}$ & 1.38$_{-3}$ & 1.85$_{-3}$ & 2.23$_{-3}$
& 2.77$_{-3}$ & 3.19$_{-3}$ & 3.28$_{-3}$ & 2.93$_{-3}$ & 2.32$_{-3}$ & 1.89$%
_{-3}$ & 1.49$_{-3}$ \\ 
-1 & Total & 7.58$_{ }$ & 6.56$_{ }$ & 5.51$_{ }$ & 4.73$_{ }$ & 4.13$_{ }$
& 3.31$_{ }$ & 2.57$_{ }$ & 1.51$_{ }$ & 1.09$_{ }$ & 7.16$_{-1}$ & 5.40$%
_{-1}$ & 3.99$_{-1}$ \\ \hline
1 & 2p & 2.15$_{ }$ & 1.71$_{ }$ & 1.36$_{ }$ & 1.13$_{ }$ & 9.65$_{-1}$ & 
7.57$_{-1}$ & 5.79$_{-1}$ & 3.36$_{-1}$ & 2.43$_{-1}$ & 1.59$_{-1}$ & 1.20$%
_{-1}$ & 8.87$_{-2}$ \\ 
1 & 2s & 6.30$_{-1}$ & 5.22$_{-1}$ & 4.09$_{-1}$ & 3.34$_{-1}$ & 2.81$_{-1}$
& 2.14$_{-1}$ & 1.59$_{-1}$ & 8.71$_{-2}$ & 6.09$_{-2}$ & 3.85$_{-2}$ & 2.85$%
_{-2}$ & 2.06$_{-2}$ \\ 
1 & 1s & 1.75$_{-4}$ & 9.47$_{-4}$ & 1.85$_{-3}$ & 2.62$_{-3}$ & 3.21$_{-3}$
& 3.93$_{-3}$ & 4.32$_{-3}$ & 3.92$_{-3}$ & 3.30$_{-3}$ & 2.46$_{-3}$ & 1.96$%
_{-3}$ & 1.52$_{-3}$ \\
\hline\hline
\end{tabular}%
\end{table}

\newpage

\begin{table}[htbp]
\caption{Total ionization cross sections for multicharge bare ions on Fluor
at different impact energies. Cross sections are in atomic units divided $%
Z_P^2$, and projectile energies are in MeV/amu}%
\begin{tabular}{|r|r|llllllllllll|}
\hline\hline
$Z_P$ & $nl$ & \ 0.1 & 0.2 & 0.3 & 0.4 & 0.5 & 0.7 & 1 & 2 & 3 & 5 & 7 & 10
\\ 
-1 & 2p & 1.11$_{ }$ & 1.00$_{ }$ & 8.68$_{-1}$ & 7.57$_{-1}$ & 6.70$_{-1}$
& 5.46$_{-1}$ & 4.29$_{-1}$ & 2.57$_{-1}$ & 1.87$_{-1}$ & 1.23$_{-1}$ & 9.35$%
_{-2}$ & 6.94$_{-2}$ \\ 
-1 & 2s & 3.66$_{-1}$ & 3.26$_{-1}$ & 2.71$_{-1}$ & 2.29$_{-1}$ & 1.98$_{-1}$
& 1.55$_{-1}$ & 1.17$_{-1}$ & 6.63$_{-2}$ & 4.68$_{-2}$ & 2.99$_{-2}$ & 2.22$%
_{-2}$ & 1.61$_{-2}$ \\ 
-1 & 1s & 1.08$_{-4}$ & 3.65$_{-4}$ & 6.49$_{-4}$ & 9.14$_{-4}$ & 1.15$_{-3}$
& 1.50$_{-3}$ & 1.84$_{-3}$ & 2.10$_{-3}$ & 1.97$_{-3}$ & 1.63$_{-3}$ & 1.36$%
_{-3}$ & 1.08$_{-3}$ \\ 
-1 & Total & 6.28$_{ }$ & 5.67$_{ }$ & 4.88$_{ }$ & 4.25$_{ }$ & 3.75$_{ }$
& 3.04$_{ }$ & 2.38$_{ }$ & 1.42$_{ }$ & 1.03$_{ }$ & 6.81$_{-1}$ & 5.15$%
_{-1}$ & 3.81$_{-1}$ \\ \hline
1 & 2p & 1.48$_{ }$ & 1.25$_{ }$ & 1.01$_{ }$ & 8.49$_{-1}$ & 7.31$_{-1}$ & 
5.77$_{-1}$ & 4.44$_{-1}$ & 2.60$_{-1}$ & 1.88$_{-1}$ & 1.24$_{-1}$ & 9.37$%
_{-2}$ & 6.95$_{-2}$ \\ 
1 & 2s & 3.91$_{-1}$ & 3.63$_{-1}$ & 2.96$_{-1}$ & 2.46$_{-1}$ & 2.10$_{-1}$
& 1.62$_{-1}$ & 1.21$_{-1}$ & 6.71$_{-2}$ & 4.71$_{-2}$ & 3.00$_{-2}$ & 2.22$%
_{-2}$ & 1.61$_{-2}$ \\ 
1 & 1s & 5.93$_{-5}$ & 3.79$_{-4}$ & 8.18$_{-4}$ & 1.24$_{-3}$ & 1.60$_{-3}$
& 2.12$_{-3}$ & 2.51$_{-3}$ & 2.57$_{-3}$ & 2.26$_{-3}$ & 1.75$_{-3}$ & 1.42$%
_{-3}$ & 1.12$_{-3}$ \\
\hline\hline
\end{tabular}%
\end{table}

\newpage

\begin{table}[htbp]
\caption{Total ionization cross sections for multicharge bare ions on Neon
at different impact energies. Cross sections are in atomic units divided $%
Z_P^2$, and projectile energies are in MeV/amu}
\label{table6}%
\begin{tabular}{|r|r|llllllllllll|}
\hline\hline
$Z_P$ & $nl$ & \ 0.1 & 0.2 & 0.3 & 0.4 & 0.5 & 0.7 & 1 & 2 & 3 & 5 & 7 & 10
\\ \hline\hline
-1 & 2p & 7.69$_{-1}$ & 7.17$_{-1}$ & 6.34$_{-1}$ & 5.61$_{-1}$ & 5.01$_{-1}$
& 4.13$_{-1}$ & 3.28$_{-1}$ & 1.99$_{-1}$ & 1.46$_{-1}$ & 9.67$_{-2}$ & 7.34$%
_{-2}$ & 5.46$_{-2}$ \\ 
-1 & 2s & 2.42$_{-1}$ & 2.31$_{-1}$ & 1.99$_{-1}$ & 1.71$_{-1}$ & 1.49$_{-1}$
& 1.19$_{-1}$ & 9.12$_{-2}$ & 5.22$_{-2}$ & 3.71$_{-2}$ & 2.38$_{-2}$ & 1.77$%
_{-2}$ & 1.29$_{-2}$ \\ 
-1 & 1s & 4.51$_{-5}$ & 1.69$_{-4}$ & 3.20$_{-4}$ & 4.71$_{-4}$ & 6.10$_{-4}$
& 8.45$_{-4}$ & 1.09$_{-3}$ & 1.37$_{-3}$ & 1.35$_{-3}$ & 1.16$_{-3}$ & 9.97$%
_{-4}$ & 8.11$_{-4}$ \\ 
-1 & Total & 5.10$_{ }$ & 4.76$_{ }$ & 4.20$_{ }$ & 3.71$_{ }$ & 3.31$_{ }$
& 2.72$_{ }$ & 2.15$_{ }$ & 1.30$_{ }$ & 9.51$_{-1}$ & 6.30$_{-1}$ & 4.78$%
_{-1}$ & 3.55$_{-1}$ \\ \hline
1 & 2p & 1.01$_{ }$ & 9.14$_{-1}$ & 7.58$_{-1}$ & 6.41$_{-1}$ & 5.56$_{-1}$
& 4.42$_{-1}$ & 3.42$_{-1}$ & 2.02$_{-1}$ & 1.47$_{-1}$ & 9.71$_{-2}$ & 7.36$%
_{-2}$ & 5.47$_{-2}$ \\ 
1 & 2s & 2.46$_{-1}$ & 2.56$_{-1}$ & 2.19$_{-1}$ & 1.86$_{-1}$ & 1.60$_{-1}$
& 1.25$_{-1}$ & 9.45$_{-2}$ & 5.30$_{-2}$ & 3.74$_{-2}$ & 2.39$_{-2}$ & 1.78$%
_{-2}$ & 1.29$_{-2}$ \\ 
1 & 1s & 2.15$_{-5}$ & 1.60$_{-4}$ & 3.78$_{-4}$ & 6.10$_{-4}$ & 8.24$_{-4}$
& 1.17$_{-3}$ & 1.49$_{-3}$ & 1.71$_{-3}$ & 1.57$_{-3}$ & 1.27$_{-3}$ & 1.06$%
_{-3}$ & 8.40$_{-4}$ \\
\hline\hline
\end{tabular}%
\end{table}

\newpage 
\begin{table}[htbp]
\caption{Total ionization cross sections for multicharge bare ions on
Phosphorous at different impact energies. Cross sections are in atomic units
divided $Z_P^2$, and projectile energies are in MeV/amu}%
\begin{tabular}{|r|r|llllllllllll|}
\hline\hline
$Z_P$ & $nl$ & \ 0.1 & 0.2 & 0.3 & 0.4 & 0.5 & 0.7 & 1 & 2 & 3 & 5 & 7 & 10
\\ \hline
-1 & 3p & 4.43$_{ }$ & 2.95$_{ }$ & 2.19$_{ }$ & 1.74$_{ }$ & 1.45$_{ }$ & 
1.08$_{ }$ & 7.95$_{-1}$ & 4.36$_{-1}$ & 3.02$_{-1}$ & 1.87$_{-1}$ & 1.39$%
_{-1}$ & 9.88$_{-2}$ \\ 
-1 & 3s & 1.67$_{ }$ & 1.06$_{ }$ & 7.70$_{-1}$ & 6.03$_{-1}$ & 4.95$_{-1}$
& 3.65$_{-1}$ & 2.63$_{-1}$ & 1.38$_{-1}$ & 9.36$_{-2}$ & 5.74$_{-2}$ & 4.22$%
_{-2}$ & 2.95$_{-2}$ \\ 
-1 & 2p & 2.04$_{-2}$ & 3.20$_{-2}$ & 3.77$_{-2}$ & 4.03$_{-2}$ & 4.12$_{-2}$
& 4.04$_{-2}$ & 3.73$_{-2}$ & 2.87$_{-2}$ & 2.21$_{-2}$ & 1.53$_{-2}$ & 1.19$%
_{-2}$ & 9.05$_{-3}$ \\ 
-1 & 2s & 9.82$_{-3}$ & 2.01$_{-2}$ & 2.44$_{-2}$ & 2.55$_{-2}$ & 2.53$_{-2}$
& 2.35$_{-2}$ & 2.04$_{-2}$ & 1.40$_{-2}$ & 1.04$_{-2}$ & 7.01$_{-3}$ & 5.35$%
_{-3}$ & 3.99$_{-3}$ \\ 
-1 & 1s & 7.52$_{-7}$ & 4.44$_{-6}$ & 1.12$_{-5}$ & 2.03$_{-5}$ & 3.09$_{-5}$
& 5.44$_{-5}$ & 9.01$_{-5}$ & 2.05$_{-4}$ & 2.50$_{-4}$ & 2.68$_{-4}$ & 2.54$%
_{-4}$ & 2.25$_{-4}$ \\ 
-1 & Total & 1.68$_{ 1}$ & 1.12$_{ 1}$ & 8.38$_{ }$ & 6.72$_{ }$ & 5.62$_{ }$
& 4.27$_{ }$ & 3.17$_{ }$ & 1.78$_{ }$ & 1.25$_{ }$ & 7.81$_{-1}$ & 5.84$%
_{-1}$ & 4.18$_{-1}$ \\ \hline
1 & 3p & 4.83$_{ }$ & 3.10$_{ }$ & 2.26$_{ }$ & 1.78$_{ }$ & 1.47$_{ }$ & 
1.10$_{ }$ & 8.00$_{-1}$ & 4.36$_{-1}$ & 3.02$_{-1}$ & 1.87$_{-1}$ & 1.39$%
_{-1}$ & 9.88$_{-2}$ \\
\hline
\end{tabular}%
\end{table}

\begin{table}[htbp]
\caption{Phosphorous Continuated}%
\begin{tabular}{|r|r|llllllllllll|}
\hline
$Z_P$ & $nl$ & \ 0.1 & 0.2 & 0.3 & 0.4 & 0.5 & 0.7 & 1 & 2 & 3 & 5 & 7 & 10
\\ \hline
6 & 3p & 8.79$_{-1}$ & 1.18$_{ }$ & 1.16$_{ }$ & 1.07$_{ }$ & 9.79$_{-1}$ & 
8.23$_{-1}$ & 6.56$_{-1}$ & 3.91$_{-1}$ & 2.80$_{-1}$ & 1.80$_{-1}$ & 1.34$%
_{-1}$ & 9.88$_{-2}$ \\ 
6 & 3s & 2.26$_{-1}$ & 3.89$_{-1}$ & 3.92$_{-1}$ & 3.64$_{-1}$ & 3.33$_{-1}$
& 2.78$_{-1}$ & 2.20$_{-1}$ & 1.27$_{-1}$ & 8.91$_{-2}$ & 5.59$_{-2}$ & 4.09$%
_{-2}$ & 2.95$_{-2}$ \\ 
6 & 2p & 2.33$_{-5}$ & 1.37$_{-3}$ & 5.61$_{-3}$ & 1.09$_{-2}$ & 1.55$_{-2}$
& 2.16$_{-2}$ & 2.51$_{-2}$ & 2.31$_{-2}$ & 1.92$_{-2}$ & 1.42$_{-2}$ & 1.13$%
_{-2}$ & 9.05$_{-3}$ \\ 
6 & 2s & 8.24$_{-6}$ & 5.32$_{-4}$ & 2.21$_{-3}$ & 4.33$_{-3}$ & 6.26$_{-3}$
& 8.95$_{-3}$ & 1.08$_{-2}$ & 1.04$_{-2}$ & 8.77$_{-3}$ & 6.43$_{-3}$ & 5.06$%
_{-3}$ & 3.99$_{-3}$ \\ 
6 & 1s & 0.00$_{ }$ & 4.38$_{-9}$ & 8.49$_{-8}$ & 4.93$_{-7}$ & 1.61$_{-6}$
& 7.29$_{-6}$ & 2.59$_{-5}$ & 1.30$_{-4}$ & 2.13$_{-4}$ & 2.74$_{-4}$ & 2.71$%
_{-4}$ & 2.25$_{-4}$ \\ 
6 & Total & 3.09$_{ }$ & 4.34$_{ }$ & 4.29$_{ }$ & 4.01$_{ }$ & 3.71$_{ }$ & 
3.17$_{ }$ & 2.58$_{ }$ & 1.59$_{ }$ & 1.15$_{ }$ & 7.50$_{-1}$ & 5.61$_{-1}$
& 4.18$_{-1}$ \\ \hline
8 & 3p & 5.40$_{-1}$ & 8.55$_{-1}$ & 8.99$_{-1}$ & 8.71$_{-1}$ & 8.22$_{-1}$
& 7.19$_{-1}$ & 5.94$_{-1}$ & 3.70$_{-1}$ & 2.69$_{-1}$ & 1.76$_{-1}$ & 1.31$%
_{-1}$ & 9.88$_{-2}$ \\
\hline\hline
\end{tabular}%
\end{table}

\newpage 
\begin{table}[htbp]
\caption{Total ionization cross sections for multicharge bare ions on Sulfur
at different impact energies. Cross sections are in atomic units divided $%
Z_P^2$, and projectile energies are in MeV/amu}%
\begin{tabular}{|r|r|llllllllllll|}
\hline\hline
$Z_P$ & $nl$ & \ 0.1 & 0.2 & 0.3 & 0.4 & 0.5 & 0.7 & 1 & 2 & 3 & 5 & 7 & 10
\\ \hline
-1 & 3p & 4.18$_{ }$ & 2.91$_{ }$ & 2.21$_{ }$ & 1.78$_{ }$ & 1.50$_{ }$ & 
1.14$_{ }$ & 8.48$_{-1}$ & 4.67$_{-1}$ & 3.28$_{-1}$ & 2.08$_{-1}$ & 1.54$%
_{-1}$ & 1.12$_{-1}$ \\ 
-1 & 3s & 1.18$_{ }$ & 7.83$_{-1}$ & 5.76$_{-1}$ & 4.55$_{-1}$ & 3.76$_{-1}$
& 2.80$_{-1}$ & 2.03$_{-1}$ & 1.07$_{-1}$ & 7.31$_{-2}$ & 4.51$_{-2}$ & 3.28$%
_{-2}$ & 2.34$_{-2}$ \\ 
-1 & 2p & 1.16$_{-2}$ & 1.94$_{-2}$ & 2.38$_{-2}$ & 2.61$_{-2}$ & 2.72$_{-2}$
& 2.76$_{-2}$ & 2.62$_{-2}$ & 2.01$_{-2}$ & 1.59$_{-2}$ & 1.14$_{-2}$ & 8.92$%
_{-3}$ & 6.81$_{-3}$ \\ 
-1 & 2s & 5.45$_{-3}$ & 1.29$_{-2}$ & 1.66$_{-2}$ & 1.81$_{-2}$ & 1.84$_{-2}$
& 1.76$_{-2}$ & 1.57$_{-2}$ & 1.08$_{-2}$ & 8.23$_{-3}$ & 5.63$_{-3}$ & 4.32$%
_{-3}$ & 3.24$_{-3}$ \\ 
-1 & 1s & 3.78$_{-7}$ & 2.37$_{-6}$ & 6.21$_{-6}$ & 1.16$_{-5}$ & 1.81$_{-5}$
& 3.30$_{-5}$ & 5.68$_{-5}$ & 1.22$_{-4}$ & 1.59$_{-4}$ & 1.86$_{-4}$ & 1.86$%
_{-4}$ & 1.72$_{-4}$ \\ 
-1 & Total & 1.50$_{ 1}$ & 1.04$_{ 1}$ & 7.96$_{ }$ & 6.46$_{ }$ & 5.45$_{ }$
& 4.19$_{ }$ & 3.14$_{ }$ & 1.76$_{ }$ & 1.24$_{ }$ & 7.95$_{-1}$ & 5.91$%
_{-1}$ & 4.31$_{-1}$ \\ \hline
1 & 3p & 4.47$_{ }$ & 3.04$_{ }$ & 2.27$_{ }$ & 1.82$_{ }$ & 1.52$_{ }$ & 
1.15$_{ }$ & 8.53$_{-1}$ & 4.68$_{-1}$ & 3.28$_{-1}$ & 2.08$_{-1}$ & 1.54$%
_{-1}$ & 1.12$_{-1}$ \\
\hline
\end{tabular}%
\end{table}

\begin{table}[htbp]
\caption{Sulfur Continuated}%
\begin{tabular}{|r|r|llllllllllll|}
\hline
$Z_P$ & $nl$ & \ 0.1 & 0.2 & 0.3 & 0.4 & 0.5 & 0.7 & 1 & 2 & 3 & 5 & 7 & 10
\\ \hline
6 & 3p & 6.65$_{-1}$ & 1.01$_{ }$ & 1.04$_{ }$ & 9.93$_{-1}$ & 9.31$_{-1}$ & 
8.07$_{-1}$ & 6.65$_{-1}$ & 4.15$_{-1}$ & 3.04$_{-1}$ & 1.99$_{-1}$ & 1.50$%
_{-1}$ & 1.10$_{-1}$ \\ 
6 & 3s & 1.17$_{-1}$ & 2.53$_{-1}$ & 2.69$_{-1}$ & 2.57$_{-1}$ & 2.39$_{-1}$
& 2.04$_{-1}$ & 1.64$_{-1}$ & 9.75$_{-2}$ & 6.91$_{-2}$ & 4.38$_{-2}$ & 3.22$%
_{-2}$ & 2.31$_{-2}$ \\ 
6 & 2p & 7.47$_{-6}$ & 5.73$_{-4}$ & 2.78$_{-3}$ & 6.00$_{-3}$ & 9.20$_{-3}$
& 1.40$_{-2}$ & 1.73$_{-2}$ & 1.70$_{-2}$ & 1.44$_{-2}$ & 1.07$_{-2}$ & 8.56$%
_{-3}$ & 6.62$_{-3}$ \\ 
6 & 2s & 3.06$_{-6}$ & 2.61$_{-4}$ & 1.26$_{-3}$ & 2.67$_{-3}$ & 4.09$_{-3}$
& 6.26$_{-3}$ & 7.94$_{-3}$ & 8.19$_{-3}$ & 7.02$_{-3}$ & 5.21$_{-3}$ & 4.12$%
_{-3}$ & 3.14$_{-3}$ \\ 
6 & 1s & 0.00$_{ }$ & 2.08$_{-9}$ & 4.23$_{-8}$ & 2.56$_{-7}$ & 8.57$_{-7}$
& 4.07$_{-6}$ & 1.53$_{-5}$ & 8.51$_{-5}$ & 1.49$_{-4}$ & 2.06$_{-4}$ & 2.12$%
_{-4}$ & 1.95$_{-4}$ \\ 
6 & Total & 2.23$_{ }$ & 3.53$_{ }$ & 3.67$_{ }$ & 3.53$_{ }$ & 3.33$_{ }$ & 
2.93$_{ }$ & 2.44$_{ }$ & 1.56$_{ }$ & 1.15$_{ }$ & 7.61$_{-1}$ & 5.74$_{-1}$
& 4.22$_{-1}$ \\ \hline
8 & 3p & 3.94$_{-1}$ & 7.05$_{-1}$ & 7.81$_{-1}$ & 7.84$_{-1}$ & 7.59$_{-1}$
& 6.89$_{-1}$ & 5.88$_{-1}$ & 3.87$_{-1}$ & 2.89$_{-1}$ & 1.94$_{-1}$ & 1.47$%
_{-1}$ & 1.08$_{-1}$ \\
\hline\hline
\end{tabular}%
\end{table}

\newpage

\begin{table}[tbph]
\caption{Total ionization cross sections for multicharge bare ions on Argon
at different impact energies. Cross sections are in atomic units divided $%
Z_{P}^{2}$, and projectile energies are in MeV/amu}%
\begin{tabular}{|r|r|llllllllllll|}
\hline\hline
$Z_P$ & $nl$ & 0.1 & 0.2 & 0.3 & 0.4 & 0.5 & 0.7 & 1 & 2 & 3 & 5 & 7 & 10 \\ 
\hline
-1 & 3p & 2.91$_{ }$ & 2.18$_{ }$ & 1.71$_{ }$ & 1.41$_{ }$ & 1.20$_{ }$ & 
9.30$_{-1}$ & 7.01$_{-1}$ & 4.11$_{-1}$ & 2.87$_{-1}$ & 1.81$_{-1}$ & 1.35$%
_{-1}$ & 9.89$_{-2}$ \\ 
-1 & 3s & 6.48$_{-1}$ & 4.70$_{-1}$ & 3.58$_{-1}$ & 2.89$_{-1}$ & 2.42$_{-1}$
& 1.83$_{-1}$ & 1.34$_{-1}$ & 7.25$_{-2}$ & 5.02$_{-2}$ & 3.13$_{-2}$ & 2.29$%
_{-2}$ & 1.65$_{-2}$ \\ 
-1 & 2p & 4.31$_{-3}$ & 7.89$_{-3}$ & 1.04$_{-2}$ & 1.20$_{-2}$ & 1.30$_{-2}$
& 1.39$_{-2}$ & 1.40$_{-2}$ & 1.17$_{-2}$ & 9.67$_{-3}$ & 7.09$_{-3}$ & 5.63$%
_{-3}$ & 4.34$_{-3}$ \\ 
-1 & 2s & 1.67$_{-3}$ & 5.29$_{-3}$ & 7.76$_{-3}$ & 9.16$_{-3}$ & 9.85$_{-3}$
& 1.01$_{-2}$ & 9.54$_{-3}$ & 7.07$_{-3}$ & 5.51$_{-3}$ & 3.84$_{-3}$ & 2.98$%
_{-3}$ & 2.25$_{-3}$ \\ 
-1 & 1s & 1.04$_{-7}$ & 7.27$_{-7}$ & 2.05$_{-6}$ & 4.05$_{-6}$ & 6.60$_{-6}$
& 1.29$_{-5}$ & 2.40$_{-5}$ & 5.96$_{-5}$ & 8.44$_{-5}$ & 1.08$_{-4}$ & 1.14$%
_{-4}$ & 1.11$_{-4}$ \\ 
-1 & Total & 1.88$_{ 1}$ & 1.41$_{ 1}$ & 1.11$_{ 1}$ & 9.11$_{ }$ & 7.77$_{
} $ & 6.05$_{ }$ & 4.58$_{ }$ & 2.70$_{ }$ & 1.89$_{ }$ & 1.20$_{ }$ & 8.97$%
_{-1}$ & 6.57$_{-1}$ \\ \hline
1 & 3p & 2.98$_{ }$ & 2.26$_{ }$ & 1.76$_{ }$ & 1.43$_{ }$ & 1.22$_{ }$ & 
9.38$_{-1}$ & 7.05$_{-1}$ & 4.12$_{-1}$ & 2.87$_{-1}$ & 1.81$_{-1}$ & 1.35$%
_{-1}$ & 9.90$_{-2}$ \\
\hline
\end{tabular}%
\end{table}

\newpage

\begin{table}[htbp]
\caption{Continuation of TABLE VII for Argon target.}%
\begin{tabular}{|r|r|llllllllllll|}
\hline
$Z_P$ & $nl$ & 0.1 & 0.2 & 0.3 & 0.4 & 0.5 & 0.7 & 1 & 2 & 3 & 5 & 7 & 10 \\ 
\hline
6 & 3p & 2.91$_{-1}$ & 5.80$_{-1}$ & 6.62$_{-1}$ & 6.72$_{-1}$ & 6.54$_{-1}$
& 5.97$_{-1}$ & 5.14$_{-1}$ & 3.41$_{-1}$ & 2.60$_{-1}$ & 1.71$_{-1}$ & 1.30$%
_{-1}$ & 9.65$_{-2}$ \\ 
6 & 3s & 3.39$_{-2}$ & 1.20$_{-1}$ & 1.44$_{-1}$ & 1.45$_{-1}$ & 1.39$_{-1}$
& 1.23$_{-1}$ & 1.03$_{-1}$ & 6.43$_{-2}$ & 4.66$_{-2}$ & 3.01$_{-2}$ & 2.23$%
_{-2}$ & 1.62$_{-2}$ \\ 
6 & 2p & 1.12$_{-6}$ & 1.21$_{-4}$ & 7.56$_{-4}$ & 1.96$_{-3}$ & 3.42$_{-3}$
& 6.15$_{-3}$ & 8.77$_{-3}$ & 1.01$_{-2}$ & 8.87$_{-3}$ & 6.77$_{-3}$ & 5.45$%
_{-3}$ & 4.25$_{-3}$ \\ 
6 & 2s & 4.85$_{-7}$ & 6.71$_{-5}$ & 4.12$_{-4}$ & 1.03$_{-3}$ & 1.76$_{-3}$
& 3.10$_{-3}$ & 4.39$_{-3}$ & 5.17$_{-3}$ & 4.63$_{-3}$ & 3.54$_{-3}$ & 2.84$%
_{-3}$ & 2.18$_{-3}$ \\ 
6 & 1s & 0.00$_{ }$ & 0.46$_{-9}$ & 1.05$_{-8}$ & 6.80$_{-8}$ & 2.45$_{-7}$
& 1.29$_{-6}$ & 5.41$_{-6}$ & 3.77$_{-5}$ & 7.46$_{-5}$ & 1.19$_{-4}$ & 1.31$%
_{-4}$ & 1.28$_{-4}$ \\ 
6 & Total & 1.81$_{ }$ & 3.72$_{ }$ & 4.27$_{ }$ & 4.33$_{ }$ & 4.23$_{ }$ & 
3.87$_{ }$ & 3.35$_{ }$ & 2.24$_{ }$ & 1.72$_{ }$ & 1.14$_{ }$ & 8.64$_{-1}$
& 6.41$_{-1}$ \\ \hline
8 & 3p & 1.55$_{-1}$ & 3.83$_{-1}$ & 4.72$_{-1}$ & 5.04$_{-1}$ & 5.10$_{-1}$
& 4.90$_{-1}$ & 4.40$_{-1}$ & 3.12$_{-1}$ & 2.44$_{-1}$ & 1.65$_{-1}$ & 1.27$%
_{-1}$ & 9.47$_{-2}$ \\
\hline\hline
\end{tabular}%
\end{table}

\end{document}